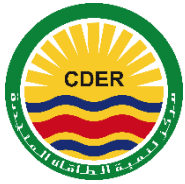

**Journal of Renewable Energies**

*Revue des Energies Renouvelables*

journal home page: https://revue.cder.dz/index.php/rer

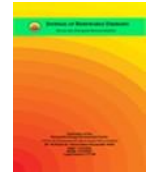

Conference paper

# Supervision of a Photovoltaic/Batteries System for Stand Alone Applications

**Djamila Rekioua** [a,*], **Saloua Belaid** [a], **Pierre Olivier Logerais** [b], **Toufik Rekioua** [a], **Zahra Mokrani** [a], **Khoudir Kakouche** [a], **Adel Oubelaid** [a], **Faika Zaouche** [a]

[a] *Université de Bejaia, Faculté de Technologie, Laboratoire de Technologie Industrielle et de l'Information, Bejaia, Algeria,*

[b] *Univ. Paris-Est, CERTES, IUT de Sénart-Fontainebleau, Lieusaint, France*

| ARTICLE INFO | ABSTRACT |
|---|---|
|  | Our paper is focused on optimal and control of an isolated photovoltaic system with batteries. The control is made by the application of a power management control (PMC). Batteries are kept safe from deep discharges and overloads by the PMC, maintaining a continuous supply to the load. The ease, with which this method can be implemented, as well as its effectiveness without imposing a large computing strain on the user, is noteworthy. The batteries and PV panels in the system under study are connected to a bidirectional converter enabling the batteries to be charged and drained in accordance with weather conditions. The simulation results, clearly highlight good performance of the proposed control across two different profiles. |

## 1. INTRODUCTION

Global warming and the depletion of fossil fuels have become urgent concerns that require changes to clean, limitless, renewable energy sources that can satisfy demand. The first objective of the presented application is to track the maximum power across vaiable conditions using an Maximum Power Point Tracking (MPPT) technique, ensuring that the photovoltaic (PV) system consistently operates at optimal efficiency. (Elgendy et al., 2012; Kamarzaman & Tan, 2014; Mohapatra et al., 2017).

Several MPPT strategies have been created in an effort to increase PV system efficiency. These include more sophisticated approaches like fuzzy logic controllers (FLC) and sliding mode controllers (SMC), as well as more traditional ones like perturbation and observation (P&O) and incremental conductance (Inc Cond) (Attia, 2018). There is a substantial body of research that focuses on different applications when discussing PV system power management. For example, it has been demonstrated that intelligent power management systems can retain efficiency in the face of demand fluctuations and climate variables (Soulatiantork et al., 2018). A fuzzy energy management strategy was presented by Rekioua and Matagne (2012) and Rekioua et al. (2009) for rural electrification.

---

* *Corresponding author, E-mail address: djamila.ziani@univ-bejaia.dz*
  *Tel : + 213 34215090*

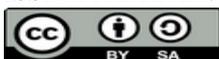

In this work, FLC is applied in power management of PV system with battery. This application is made to manage erratic changes in load needs and weather. The system consists of batteries and PV generator that both provide power to a load and FLC with P&O methods are applied to extract maximum power. Our suggested power management system satisfies energy needs, safeguards the batteries from deep discharge and overcharging, and maximizes power output from the PV generator. According to the results under Matlab/Simulink, the proposed power management system is effectively developed.

## 2. PROPOSED SYSTEM

The studied system is shown in Fig.1. Three switches are necessary for power management: $K_1$, $K_2$, and $K_3$.

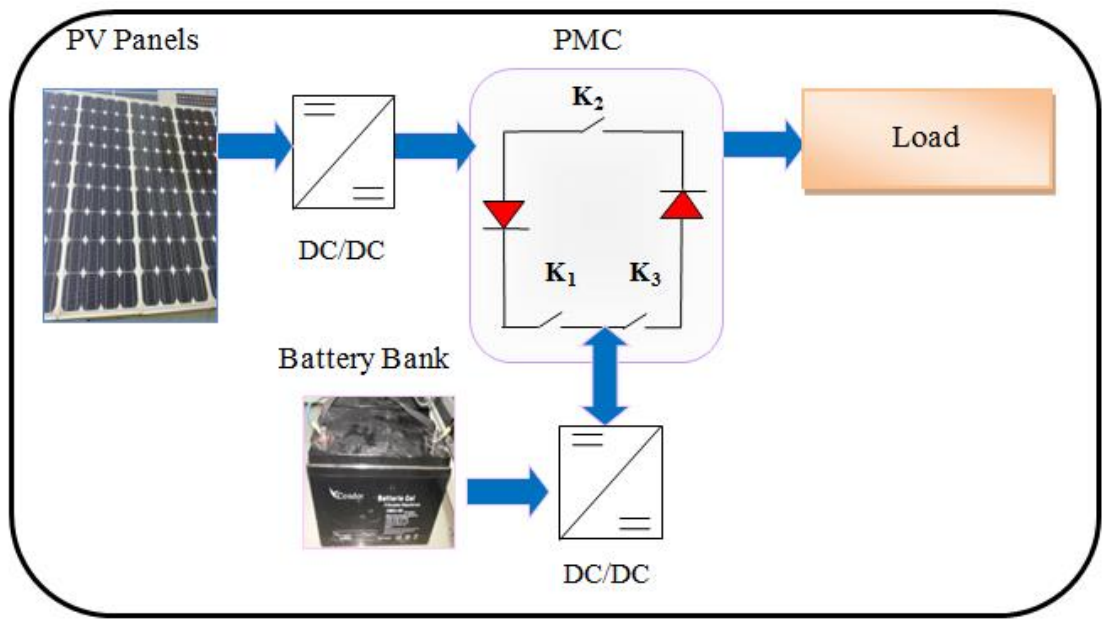

Fig.1. Proposed studied system

### 2.1 Photovoltaic panels modeling

The behavior and operation of PV systems are described by a variety of mathematical models. We explore the model shown in Fig. 2 in this work (Bratcu et al., 2008; Dursun & Kilic, 2012; Naveen Kumar, 2014). The following equations determine this model's $I_{pv}(V_{pv})$ and $I_{pv}(V_{pv})$ characteristic:

$$I_{pv} = I_{ph} - I_d - I_{Rsh} \qquad (1)$$

$$I_{pv} = I_{ph} - I_0 \left[ exp\left(\frac{q(V_{pv} + R_s.I_{pv})}{A.N_s.K.T_j}\right) \right] - \frac{V_{pv} + R_s.I_{pv}}{R_{sh}} \qquad (2)$$

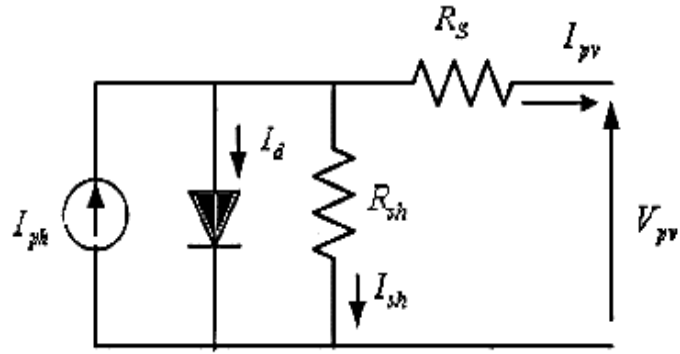

Electrical characteristics obtained used the PV panels are shown in Fig.3.

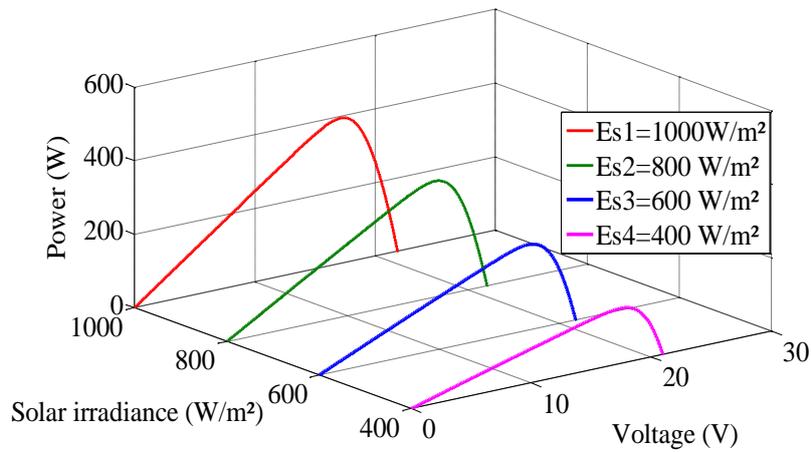

Fig 3. PV electrical characteristics

## 2.2 Battery model

Figure 4 displays the model that was used for this paper. A voltage source and an internal resistance are its two electrical components (Hajizadeh & Golkar, 2007, Singh & Snehlata, 2011; Rekioua, 2023).

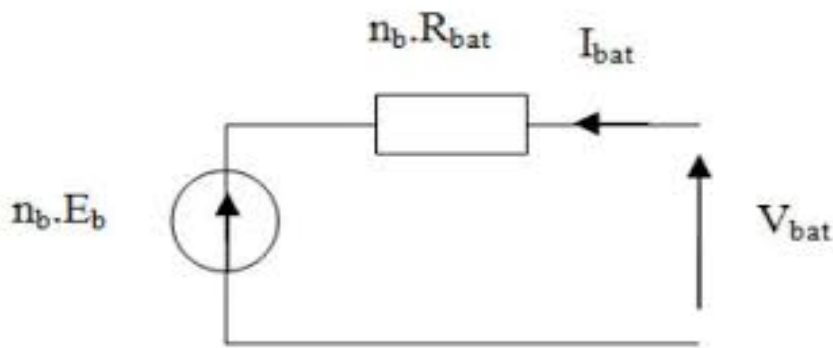

Fig. 4 Battery electrical model

$$I_{bat} = E_b \pm n_{bat.} R_{bat}. I_{bat} \tag{3}$$

The battery capacity C$_{bat}$ is defined as (Kamarzaman & Tan, 2014):

$$C_{bat} = C_{10} \frac{1.76(1 + 0.005\Delta T)}{1 + 0.67\left(\frac{I_{bat}}{I_{10}}\right)} R_{bat}.I_{bat} \tag{4}$$

The following represents the battery's state of charge:

$$SOC = 1 - \frac{Q}{C_{bat}} \tag{5}$$

$$Q = I_{bat}.t \tag{6}$$

The charge and discharge voltage are given as:

$$\begin{aligned}V_{bat-dis} = {}& n_{Bat-serial}(1.965 + 0.12\ SOC) \\ & - n_{Bat-serial}\frac{|I_{bat}|}{C_{10}} x \left(\frac{4}{1 + (|I_{bat}|)^{1,8}} + \frac{0.27}{SOC^{1.5}} + 0.02\right).(1 - 0.007\Delta T)\end{aligned} \tag{7}$$

$$\begin{aligned}V_{bat-ch} = {}& n_{Bat-serial}(2 + 0.16\ SOC) \\ & + n_{Bat-serial}\frac{|I_{bat}|}{C_{10}} \left(\frac{6}{1 + (|I_{bat}|)^{0,86}} + \frac{0.48}{SOC^{1.2}} + 0.036\right)(1 - 0.025\Delta T)\end{aligned} \tag{8}$$

## 3. MPPT CONTROL

### 3.1. Perturb & Observe method

Among the most popular traditional techniques is the P&O method (Idjdarene et al; 2011). It's algorithm's flowchart is given in Fig. 5. The converter output voltage and current are determined as (Atia & al., 2012; Rekioua & al., 2014; Tamalouzt & al., 2016):

$$V_{out} = V_{pv}\left(\frac{1}{1-D}\right) \tag{9}$$

$$I_{out} = I_{pv}(1-D) \tag{10}$$

### 3.2 Fuzzy logic control (FLC) method

The error (E) and change in error (CE) are the two inputs that make up the MPPT fuzzy logic controller system.

$$E(k) = \frac{P(k) - P(k-1)}{V(k) - V(k-1)} \tag{11}$$

$$CE(k) = E(k) - E(k-1) \tag{12}$$

Table 1 shows the different rules.

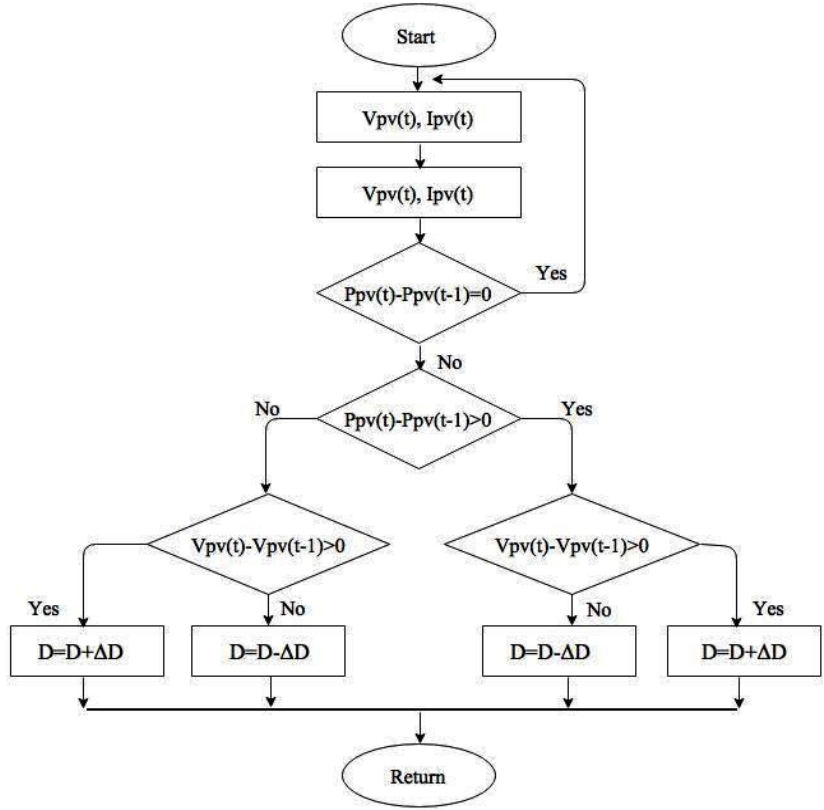

Fig.5 P&O flowchart

Table 1. Rule Base Table

| CE | E | NB | NS | Z | PS | PB |
|---|---|---|---|---|---|---|
| NB | | NB | NB | NS | NS | Z |
| NS | | NB | NS | NS | Z | PS |
| Z | | NS | NS | Z | PS | PS |
| PS | | NS | Z | PS | PS | PB |
| PB | | Z | PS | PS | PB | PB |

## 3.3. Comparison between MPPT's results

By comparing the two MPPT methods under the same conditions, it is clear that FLC responses rapidly and is more precise than the P&O which presents oscillations (Fig.6).

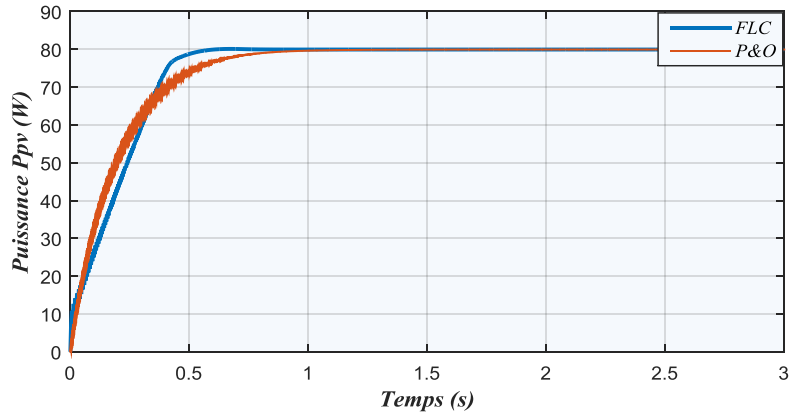

Fig.6. Photovoltaic power waveform

## 4. POWER MANAGEMENT CONTROL

According to different works of Mokrani et al. (2014), Mebarki et al. (2015), Mohammedi et al. (2016), Mebarki et al. (2016), Zaouche et al. (2017), Rekioua (2018), Khiareddine et al. (2018), Rahrah et al. (2015), Belaid et al. (2022), Kakouche et al. (2022), Mohamed et al (2023), PMC permits the PV generator to run at maximum power, protects batteries, and fulfils the need for energy. The three switch states ($K_1$, $K_2$, and $K_3$) determine the suggested control. The operation system with five modes modes, depends on the various tests (Fig.7). The various modes are obtained as (Table 2).

Table 2.Various modes

| Modes | $K_1$ | $K_2$ | $K_3$ |
|---|---|---|---|
| Mode1 | On | On | Off |
| Mode2 | Off | On | On |
| Mode3 | Off | Off | On |
| Mode4 | Off | On | Off |
| Mode5 | Off | Off | Off |

Mode 1: PV power $P_{PV}$ is more than enough to fed the load and refill batteries.

Mode 2: PV power is insufficient ($0<P_{pv}<P_{load}$), so battery power is added to supply the load.

Mode 3: Only the batteries can supply the load ($P_{av}<0$).

Mode 4: Disconnecting batteries is required for protection.

Mode 5: Batteries are depleted and the PV generator is not producing in this mode. The load has been cut off.

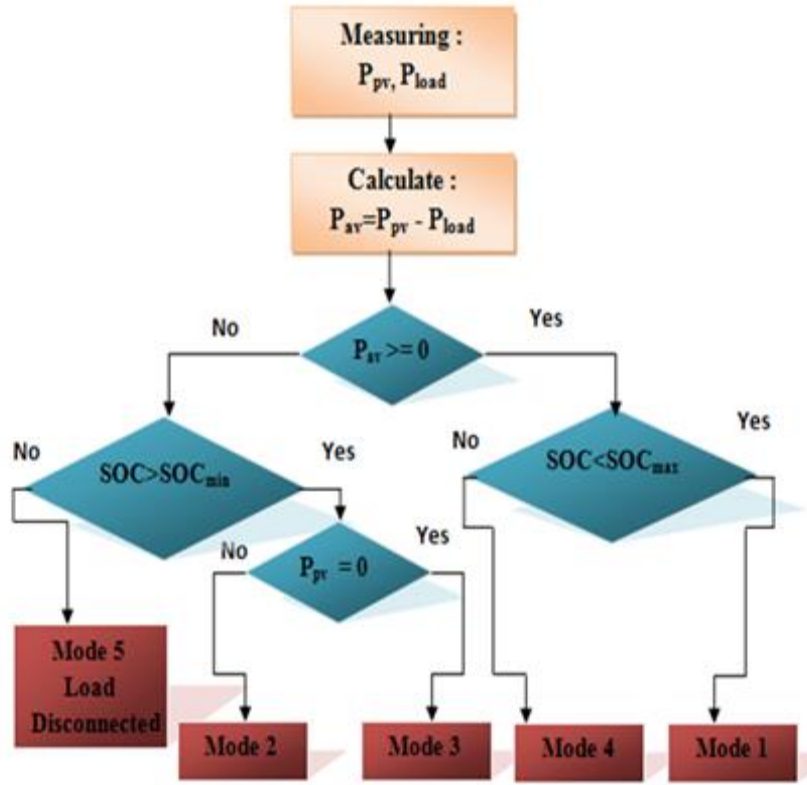

Fig. 7. PMC flowchart

The graphic below (Fig.8) depicts the average consumption profile:

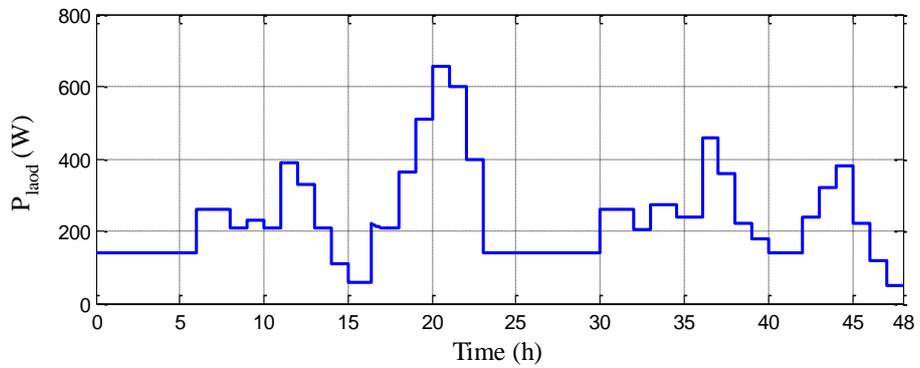

Fig. 8. Load Profile

The profile of temperature and irradiation is given as (Fig.9) :

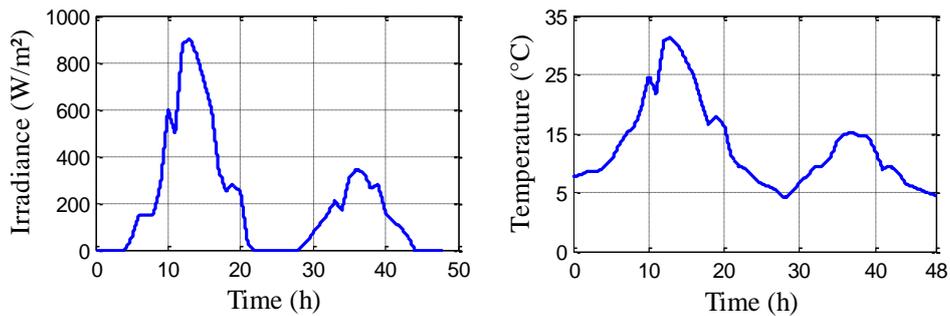

Fig. 9. Profile of temperature and irradance

Next, the various powers develop as (Fig.10):

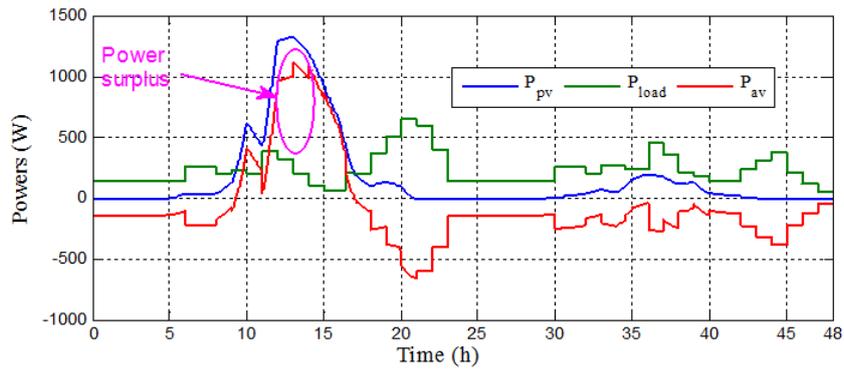

Fig. 10. Various developed powers

## 5. CONCLUSION

This paper presents the battery-storage photovoltaic system's power management control. The primary component characteristics and the system setup are provided. Verification of the system performances has been done using simulation results for the application. The findings produced demonstrate how well the chosen control approach worked. The proposed control permits to maintain good performance, whatever the weather variations. It also guarantees that the suggested application operates exactly as intended.

## NOMENCLATURE

| | | | |
|---|---|---|---|
| $C_{bat}$ | Battery capacity, [Ah] | $R_{bat}$ | Internal resistance, [Ω] |
| $C_{10}$ | Rated capacity, [Ah] | t | current the discharging time, [h] |
| E | Error | $V_{out}$ | converter's output voltage, [V] |
| $E_b$ | Voltage source, [V] | $V_{pv}$ | Photovoltaic voltage, [V] |
| $I_d$ | Diode current, [A] | CE | Change in error |
| $I_{ph}$ | Photo current, [A] | DC | Direct current |
| $I_{pv}$ | Photovoltaic current, [A] | FLC | Fuzzy logic control |
| $I_{Rsh}$ | Shunt current, [A] | MPPT | Maximum power point tracking |
| $K_i (i=1,2,3)$ | Switches | PMC | Power management control |
| $R_{sh}$ | Shunt resistance, [Ω] | PV | Photovoltaic |
| $R_s$ | Series resistance, [Ω] | P&O | Perturb & Observe |
| $I_{out}$ | Converter's output current [A] | □T | Accumulator's heat, °C |